\documentclass[12pt]{article}
\usepackage{graphicx}
\begin{document} 

\title{Advertising effects in Sznajd marketing model}

\author{Christian Schulze\\
Institute for Theoretical Physics, Cologne University\\D-50923 K\"oln, Euroland}

\maketitle
\centerline{e-mail: ab127@uni-koeln.de}

\bigskip
Abstract: The traditional Sznajd model, as well as its Ochrombel 
simplification for opinion spreading, are applied to marketing
with the help of advertising. The larger the lattice is the
smaller is the amount of advertising needed to convince the whole
market.

\bigskip


\bigskip
Science models like percolation have been applied to marketing by
word-of-mouth \cite{goljan} as well as advertising through mass
media
\cite{proykova}. The present work applies the Sznajd model of
consensus building \cite{sznajd} (see \cite{stauffer} for a 
review) to the same problem. How strong has the advertising to
be in order to help one of two products to win the whole market
even though initially this product is in the minority ?

In the Sznajd model as used here, initially a random fraction $p$ 
of the sites of a square lattice are customers of product
A, while the remaining fraction $1-p$ of customers buy product
B. At every iteration (random sequential updating) two 
neighbouring A sites convince their six neighbours to become also
A sites. In the Ochrombel simplification, one A site suffices
to convince its four neighbours \cite{ochrombel}. 
After sufficiently many iterations in a finite lattice, all 
customers have settled onto one product. In the Sznajd model
this product is the one which initially had a slight majority of 
customers, i.e. product A for $p > 1/2$ and product B for 
$p < 1/2$; for the Ochrombel modification a fraction $p$ of 
lattices end up with only A sites, the others with only B sites.
 
\begin{figure}[hbt]
\begin{center}
\end{center}
\caption{Number of successes, if advertising is switched on only after $t_1$
iterations, for 8 iterations. The different crosses refer to $t_2-t_1=2,4,8,16$
and 32, which hardly seems to matter. 
}
\end{figure}

Advertising is now included by assuming that at each iteration
every site becomes an A site with probability $\epsilon$. We 
define a success as meaning that all sites buy A, and a failure
as meaning that inspite of the advertising all sites buy B for
at least one iteration.

\begin{figure}[hbt]
\begin{center}
\end{center}
\caption{Number of successes in Ochrombel simplification with diffusion. 
From left to right the system size increases.
}
\end{figure}

Diffusion makes the model more realistic by assuming 
\cite{schelling} that only half of the sites are occupied; at 
each iteration each agent (= occupied site) moves into
a randomly selected neighbour site if that neighbour is empty.
(``Diffusion'' is used here in the physics sense, not in that
of Bass marketing theory \cite{goljan}).

Feedback takes into account that advertising is diminished for
already successful products. The fraction of A customers is 
called $x$, that of B customers is $y=1-x$. Then advertising 
produces an A site no longer with probability $\epsilon$ but
with probability
$$ \epsilon y(t)/y(t=0)$$
at iteration $t$.

\begin{figure}[hbt]
\begin{center}
\end{center}
\caption{Log-log plot of the level of advertising needed to convert half the 
failures to successes, versus linear lattice dimension $L$. The line has the
slope -2.3.
}
\end{figure}

This defines our advertising model which we now simulate with and
without feedback, using the original Sznajd version, the 
Ochrombel simplification, and for Ochrombel also
with diffusion. We used 1000 different samples for each lattice 
size $L \times L, \; L = 31, 53, 71, 101, 301$. The initial
fraction of A agents was always $p = 0.4$.

For the Sznajd model we found that without advertising nearly all
samples ended with product B. With $\epsilon \sim 0.1$ and 
larger, all samples were successes, i.e. the advertising 
convinced 
everybody even though initially only a 40 percent minority
was convinced. Equal numbers of failures and successes were found
for $\epsilon \sim$ 0.04 and 0.025 for $L$ = 31 and 53 (not
shown). Thus 
already a small fraction of advertising is sufficient to change
nearly all samples from product B to product A. 

We also simulated ``ageing'' by having advertising only for
$t_1 < t < t_2$ Strong advertising $\epsilon = 0.5$ produced
nearly always a success for $t_1$ up to 10 and was quite useless
for $t_1 > 100$ at $L=31$, Fig.1; for larger lattices the
characteristic times are larger. The difference $t_2-t_1 \sim 10$
was less important. In short, if a mass media campaign starts
too late, then word-of-mouth propaganda through the standard 
Sznajd process has already cornered the market.

The Ochrombel simplification, that already a single site 
convinces its neighbours, is numerically much easier since no
critical point at $p=1/2$ occurs without advertising. Thus far 
less iterations are needed when failures and successes are
nearly balanced. Without advertising we have 400 A fixed points
and 600 B fixed points, and thus we ask how much advertising is
needed to reduce the number of failures from 600 to 300. Fig.2
includes diffusion and
shows that the needed $\epsilon$ increases from 0.0001 to 0.001
if $L$ decreases from 101 to 31; the transition curves all have
roughly the same shape. Without diffusion the needed advertising is slightly
larger. Fig.3 shows that the needed advertising
decreases roughly as $1/L^{2.3}$. 

Quite similar results are obtained also with feedback (for both Sznajd and 
Ochrombel version), i.e. the model's 
results are quite robust. Again, the Ochrombel version requires
less advertising than the original Sznajd version.
 
In summary, the Sznajd model and in particular its Ochrombel 
simplification are suitable to show successes and failures
of advertising to convince a market.

We thank Deutsche Forschungsgemeinschaft for support, W. Selke 
for suggesting this work, and D. Stauffer for help.

\bigskip
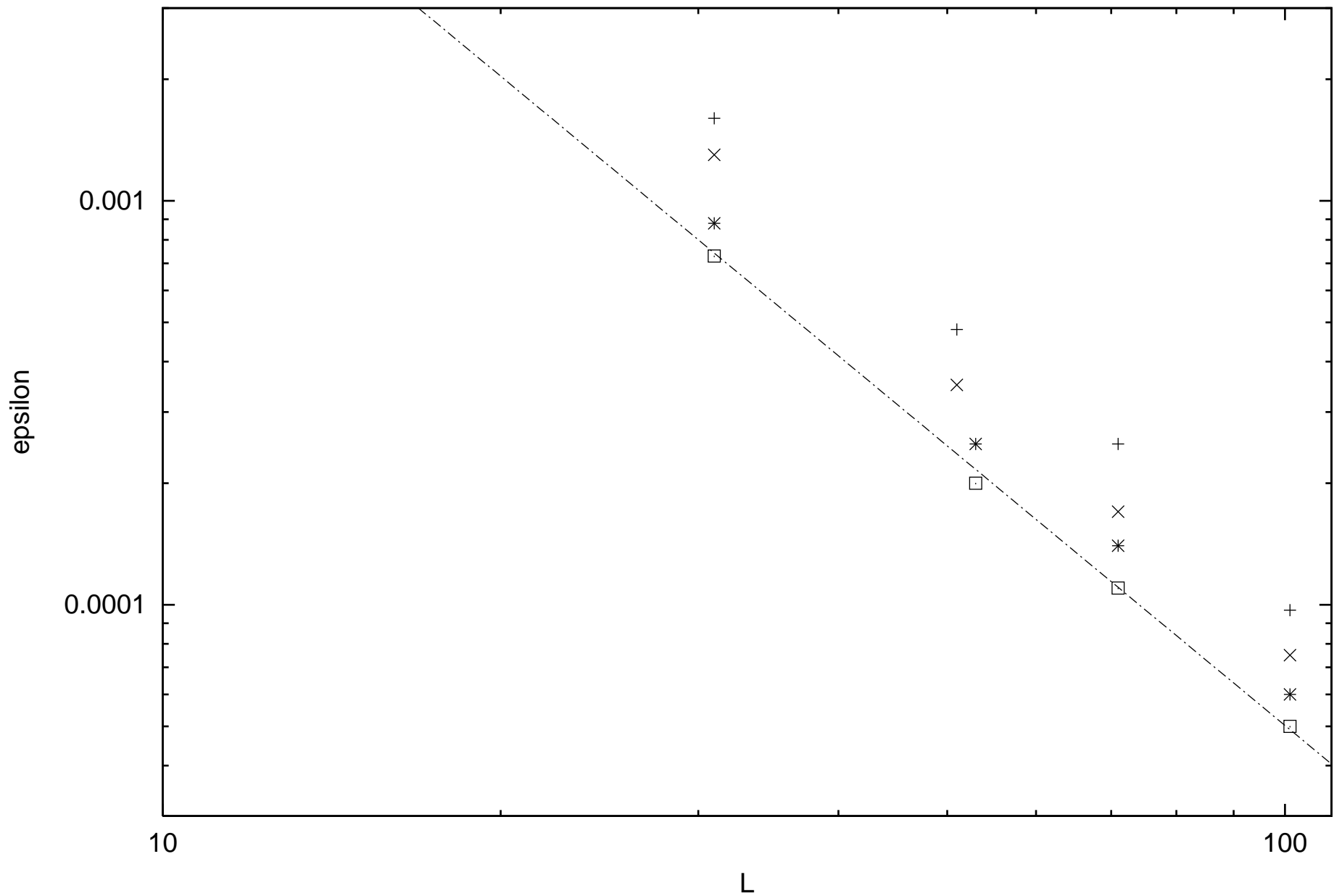
\end{document}